\documentclass[twocolumn,aps,prl,superscriptaddress,showpacs,floatfix,longbibliography]{revtex4-1}
\usepackage{url}
\usepackage{cancel}
\usepackage[colorlinks,linkcolor=blue,citecolor=blue,filecolor=black,urlcolor=blue]{hyperref}
\usepackage{epsfig,graphics}
\usepackage{graphicx}
\usepackage{dcolumn}
\usepackage{bm}
\usepackage[usenames]{color}
\usepackage{amssymb}
\usepackage{amsmath}
\usepackage{multirow}
\usepackage{float}
\usepackage{harpoon}
\usepackage{MnSymbol}
\usepackage{appendix}
\usepackage{color}
\usepackage{hyperref}
\usepackage{cleveref}
\usepackage{physics}
\usepackage{natbib}
\usepackage{bm}
\usepackage{orcidlink}

\begin{document}

\date{\today}

\title{Measuring short-range correlations using relativistic heavy-ion collisions}

\author{Lu-Meng Liu \orcidlink{0000-0001-5243-5549}}\email{liulumeng@fudan.edu.cn}
\affiliation{Physics Department and Center for Particle Physics and Field Theory, Fudan University, Shanghai 200438, China}

\author{Jun Xu}\email[Correspond to\ ]{junxu@tongji.edu.cn}
\affiliation{School of Physics Science and Engineering, Tongji University, Shanghai 200092, China}
\affiliation{Southern Center for Nuclear-Science Theory (SCNT), Institute of Modern Physics, Chinese Academy of Sciences, Huizhou 516000, Guangdong Province, China}


\author{Xu-Guang Huang}\email{huangxuguang@fudan.edu.cn}
\affiliation{Physics Department and Center for Particle Physics and Field Theory, Fudan University, Shanghai 200438, China}
\affiliation{Key Laboratory of Nuclear Physics and Ion-beam Application (MOE), Fudan University, Shanghai 200433, China}
\affiliation{Shanghai Research Center for Theoretical Nuclear Physics, National Natural Science Foundation of China and Fudan University, Shanghai 200438, China}

\begin{abstract}
We propose a novel method to measure the strength and isospin dependence of short-range correlations (SRCs) in nuclei with their collisions at relativistic energies. Since nucleons in the high-momentum tail (HMT) induced by SRCs have large transverse velocities, we find that measuring the yield ratio of free spectator neutrons at large to small transverse distances detected by Zero-Degree Calorimeter (ZDC) array may probe the fraction of nucleons in the HMT. In addition, measuring the yield ratio of ZDC neutrons in central $^{96}\mathrm{Zr}+^{96}\mathrm{Zr}$ to $^{96}\mathrm{Ru}+^{96}\mathrm{Ru}$ collisions at large transverse distances may probe the isospin dependence of the HMT and SRCs. Compared to traditional scattering experiments, our study illustrates that relativistic heavy-ion collisions may serve as an alternative way of measuring SRCs in colliding nuclei with less final-state interactions (FSIs).
\end{abstract}
\maketitle

{\it Introduction.}
Understanding fundamental nuclear forces and nucleus structures at high precision remains a major challenge of nuclear physics. The tensor force, predominantly mediated by pion exchange between nucleons, enhances spin-triplet isospin-singlet neutron-proton ($np$) SRCs at relative momenta between 300 and 600 MeV/c~\cite{Schiavilla:2006xx,Carlson:2014vla}, overwhelming proton-proton ($pp$) or neutron-neutron ($nn$) correlations in symmetric nuclear systems~\cite{Piasetzky:2006ai,Cruz-Torres:2019fum}. The correlated nucleons have larger relative momenta compared to their smaller center-of-mass momenta, leading to HMTs in one-body nucleon momentum distribution $n(k)$ above the Fermi momentum scaling as \( k^{-4} \) in both nuclear matter and finite nuclei~\cite{Subedi:2008zz,Hen:2014nza,Hen:2014lia,Hen:2016kwk}. This phenomenon may originate from quark-level properties~\cite{Hen:2016kwk,nCTEQ:2023cpo} and is related to the EMC effect~\cite{EuropeanMuon:1983wih,Weinstein:2010rt}.

Theoretically, ab initio few-body calculations have become benchmarks in quantifying such effects. Quantum Monte Carlo methods achieve high precision in light nuclei using realistic nuclear potentials and those derived from chiral effective field theory~\cite{CiofidegliAtti:1995qe,Wiringa:2013ala,Piarulli:2022ulk,Carlson:2014vla}. Complementary theoretical approaches like the no-core shell model~\cite{Barrett:2013nh} and nuclear lattice effective field theory~\cite{Freer:2017gip} provide beyond mean-field calculations including SRCs for nuclei with mass number from 2 to 40. For heavy nuclei, the description of the HMT induced by SRCs often relies on approximated methods~\cite{Weiss:2015mba,Cruz-Torres:2017sjy,Colle:2015ena,Tropiano:2021qgf} or parametrization~\cite{Cai:2016ewl}. In studies on nuclear matter, the SRC has been found to affect the relative contributions from the kinetic and potential part to the nuclear symmetry energy~\cite{Hen:2014yfa} as well as properties of neutron-star matter~\cite{Cai:2015xga} (see Ref.~\cite{Cai:2025txx} for a recent review).

Experimentally, electron-nucleus scatterings are generally used to measure the HMT induced by SRCs in finite nuclei~\cite{Subedi:2008zz,Hen:2014nza,Hen:2016kwk}. A scaling factor \( a_2(A) \), defined as the ratio of the scaled nucleon numbers in the HMT for a nucleus to that for deuteron, is used to quantify the strength of the SRC~\cite{Hen:2014lia}. Empirical studies have shown that \( a_2(A) \) increases with \( A \), saturates at about 5 for nuclei with mass number \( A \gtrsim 12 \)~\cite{Fomin:2011ng,CLAS:2019vsb,Arrington:2022sov}, and is extrapolated to \( a_2(\infty) = 7 \pm 1 \) in normal nuclear matter~\cite{CiofidegliAtti:1990rw}. 
The electron-nucleus scattering experiments have also measured the isospin dependence of the SRC, and a constant high-momentum ratio of yields from $(e,e'n)$ to $(e,e'p)$ scatterings is observed from $^{12}$C to $^{208}$Pb~\cite{CLAS:2018yvt}, indicating approximately equal numbers of protons and neutrons in the HMT. This finding shows that the $np$ SRC dominates in medium to heavy nuclei, with the ratio of $np$ to $pp$ correlations of about 20, while this ratio is reduced to about 4 in \(^4\mathrm{He} \) and 2 in \( ^3\mathrm{He} \) and \( ^3\mathrm{H} \)~\cite{Li:2022fhh}. Since electron-nucleus scattering experiments suffer from long interacting time and strong FSIs, knock-out reactions with inverse kinematics by selecting particular products in order to suppress FSIs have been performed~\cite{BMN:2021vhy}. Accurate measurement of $a_2(A)$ as well as isospin properties of the HMT may help to understand the SRC originated from nuclear forces at high precision.

The present study proposes a novel approach for measuring properties of SRCs in nuclei using free spectator nucleons at forward rapidities in relativistic heavy-ion collisions, free from the complicated midrapidity dynamics, and with extremely short interacting time ($< 10^{-24}$ s) and thus less FSIs. The projection of the Fermi motion of spectator nucleons in colliding nuclei on the transverse plane can be extracted through the spatial distribution in the coordinate space measured by ZDCs. Analyzing the spatial distribution with available high-statistics experimental data may help to extract detailed information of SRCs, including the strength and isospin dependence.



{\it Parametrization of HMT in heavy nuclei.}
Ideally, the phase-space distribution function $f_J(\mathbf{r}, \mathbf{k})$, representing the probability for a nucleon with isospin $J$ to appear at position $\mathbf{r}$ and momentum $\mathbf{k}$ in a nucleus, should be obtained from quantum many-body calculations. By slightly modifying the form in Ref.~\cite{Cai:2016ewl}, we parametrize the phase-space distribution functions for neutrons and protons in nuclei as
\begin{equation}
f_J(\mathbf{r}, \mathbf{k}) = 2 \times
\begin{cases}
\Delta_J, & 0 < |\mathbf{k}| < k_0, \\
C \left[ \frac{k_0(\mathbf{r})}{\mathbf{k}} \right]^4, & k_0 < |\mathbf{k}| < \phi k_0.
\end{cases}
\label{eq:momentum_distribution}
\end{equation}
In the above, the factor 2 accounts for the spin degeneracy. $k_0(\mathbf{r}) = \alpha_J k_F^J(\mathbf{r})$ is the transition momentum, where $k_F^J(\mathbf{r}) = [3\pi^2 \rho_J(\mathbf{r})]^{1/3}$ is the local Fermi momentum with $\rho_J(\mathbf{r})$ being the local density for nucleons with isospin $J$. Here $\alpha_J = \alpha_0(1+\alpha_1 \tau_3^J \delta)^{1/3}$ is a factor accounting for the difference between $k_0$ and $k_F^J$, with $\tau_3^n=+1$, $\tau_3^p = -1$, $\delta = (\rho_n-\rho_p)/(\rho_n+\rho_p)$ being the local isospin asymmetry, and $\alpha_0$ and $\alpha_1$ being parameters characterizing the shape and isospin dependence of the HMT. $\alpha_1<0$ corresponds to a larger (smaller) fraction of protons (neutrons) in the HMT in neutron-rich nuclei~\cite{CLAS:2018yvt}. $\Delta_J<1$ is a factor for the momentum distribution below the transition momentum, while that above $k_0$ follows the empirical $k^{-4}$ scaling~\cite{Subedi:2008zz,Hen:2014nza,Hen:2014lia,Hen:2016kwk}, with the parameter $C$ controlling the experimentally measured per-nucleon probability of finding a high-momentum nucleon, and $\phi\sim 2-3$ being the high-momentum cutoff representing the upper limit for the $k^{-4}$ scaling~\cite{Hen:2014lia}. 
 
The value of $\Delta_J$ can be determined from the normalization condition
\begin{equation}
\int f_J(\mathbf{r}, \mathbf{k}) \, \frac{d^3 {\mathbf{k}}}{(2\pi)^3} = \rho_J(\mathbf{r}).
\label{eq:norm_cond}
\end{equation}
Default values of $C$, $\phi$, $\alpha_0$, and $\alpha_1$ are listed in Table~\ref{tab:parameters}, which reproduce the HMT fractions of $32\%$ in symmetric nuclear matter and  $3\%$ in pure neutron matter at the saturation density. With the nucleon density distributions for different nuclei obtained from spherical Skyrme-Hartree-Fock (SHF) calculations~\cite{reinhard1991skyrme} by using the MSL0 force~\cite{Chen:2010qx}, the one-body momentum distributions can then be calculated from
\begin{equation}\label{nk}
n_J(k) = \int f_J(\mathbf{r}, \mathbf{k}) d^3\mathbf{r}.
\end{equation}
The number of nucleons in the HMT is
\begin{equation}\label{nk}
\tilde{N}_J=\frac{1}{2\pi^2}\int_{k_l}^{k_u} n_J(k) k^2dk,
\end{equation}
where $k_l$ and $k_u$ are chosen to be 300 and 600~\text{MeV/c}, respectively, according to Ref.~\cite{Hen:2014nza}.

\begin{table}[htpb]
    \centering
    \caption{Default parameters for nucleon momentum distributions in Eq.~(\ref{eq:momentum_distribution}).}
    \label{tab:parameters}
    \renewcommand{\arraystretch}{1.5}
    \setlength{\tabcolsep}{3mm}
    \vspace{0.3cm}
    \begin{tabular}{cccc}
        \hline\hline
        $C$     &   $\phi$      &   $\alpha_0$      &  $\alpha_1$  \\
        \hline 
        0.12     &  2.75      &   1.125      &  -0.9  \\
        \hline\hline 
    \end{tabular}
\end{table}

The total number of two-nucleon pairs induced by SRCs can be expressed as~\cite{JeffersonLabHallA:2020wrr, Li:2022fhh}
\begin{align}\label{eq:SRCpairNum}
   N_{np} &= N \cdot Z \cdot f_{sr}(A) \cdot p_{np}, \\
   N_{pp} &= \frac{Z\cdot(Z-1)}{2} \cdot f_{sr}(A) \cdot p_{pp}, \\
   N_{nn} &= \frac{N\cdot(N-1)}{2} \cdot f_{sr}(A) \cdot p_{pp},
\end{align}
where $N$, $Z$, and $A$ are respectively the neutron number, proton number, and mass number of a nucleus. $f_{sr}(A)$ represents the probability that two nucleons are close enough to interact via the short-range nucleon-nucleon (NN) force, and it is assumed to be the same for $np$, $pp$, and $nn$ pairs. The quantity $p^{}_{NN}$ is the probability that the NN interaction generates a high-momentum pair. While $p^{}_{NN}$ may depend on the nucleon momentum range, it is observed that the ratio $p_{np}/p_{nn}$ weakly depends on momentum~\cite{LabHallA:2014wqo}. 

The total number of nucleons in the HMT can be approximated by the sum of two-nucleon SRC pairs~\cite{Weiss:2015mba}
\begin{align}
   \tilde{N}_n &\approx  N_{np} + 2N_{nn} \nonumber \\
   &= \left[ N \cdot Z \cdot p_{np} + N \cdot (N-1) \cdot p_{pp} \right] \cdot f_{sr}(A), \label{eq:nhmt}\\
   \tilde{N}_p &\approx N_{np} + 2N_{pp} \nonumber \\
   &= \left[ N \cdot Z \cdot p_{np} + Z \cdot (Z-1) \cdot p_{pp} \right] \cdot f_{sr}(A). \label{eq:phmt}
\end{align}
The scaling factor for HMTs in the nucleus can then be expressed as
\begin{equation}\label{a2}
   a_2(A) \approx \frac{2}{A} \frac{\tilde{N}_n(A) + \tilde{N}_p(A)}{2\tilde{N}_p(d)}.
\end{equation}
Taking the ratio of Eq.~(\ref{eq:nhmt}) to Eq.~(\ref{eq:phmt}) leads to 
\begin{align}\label{eq:npppratio}
   \frac{p_{np}}{p_{pp}}  \approx \frac{ N \cdot (N-1) - \frac{\tilde{N}_n}{\tilde{N}_p} \cdot Z \cdot (Z-1)}{\frac{\tilde{N}_n}{\tilde{N}_p} \cdot N \cdot Z -  N \cdot Z},
\end{align}
which can be used to compare with that extracted from the experimental data~\cite{CLAS:2018xvc,JeffersonLabHallA:2020wrr,Li:2022fhh}.



\begin{figure}[htb]
\centering
\includegraphics[width=1\linewidth]{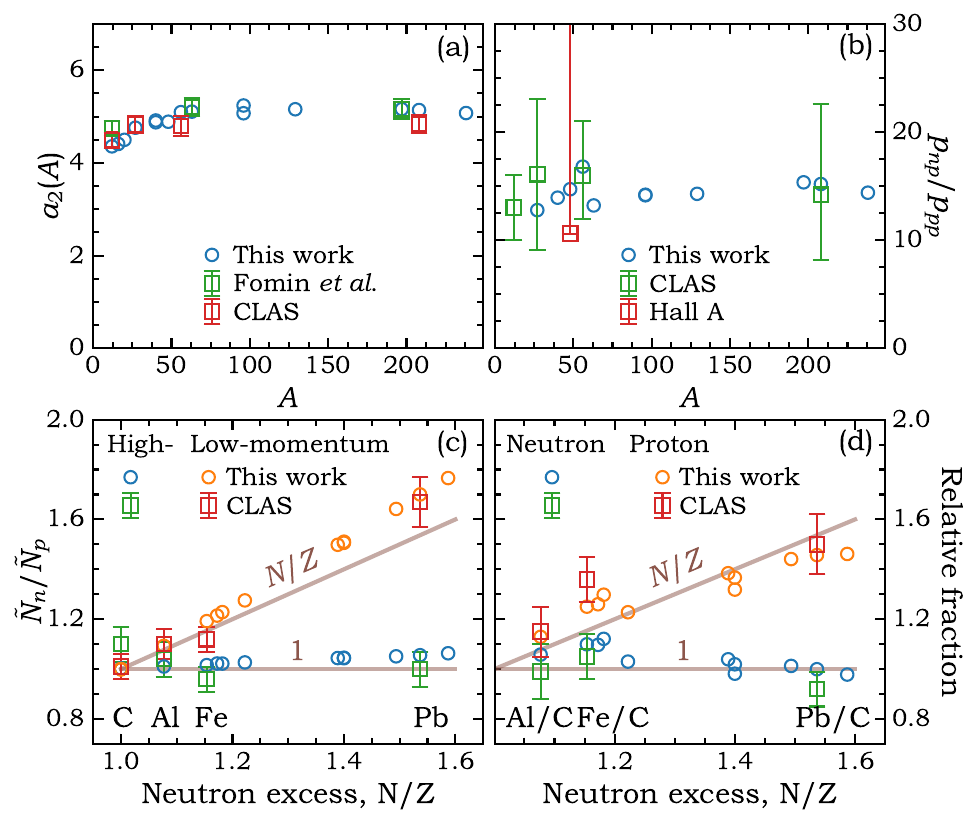}
\vspace{-0.8cm}
\caption{Upper: Comparison of the scaling factor $a_2(A)$ (a) and the ratio of $np$ to $pp$ pairing probability (b) from our model with those from electron-nucleus scattering experiments for different nuclei. Lower: Comparison of the ratio of neutron to proton numbers in high- and low-momentum regions (c) and the ratio of HMT fraction for neutrons and protons in heavy nuclei to that in $^{12}$C (d) from our model with those from electron-nucleus scattering experiments for different nuclei. The nuclei included in the comparison are $^{12}$C, $^{16}$O, $^{20}$Ne, $^{27}$Al, $^{40}$Ar, $^{40}$Ca, $^{48}$Ca, $^{56}$Fe, $^{63}$Cu, $^{96}$Zr, $^{96}$Ru, $^{129}$Xe, $^{197}$Au, $^{208}$Pb, and $^{238}$U. 
Experimental data from electron-nucleus scatterings are taken from Refs.~\cite{Fomin:2011ng,Hen:2012fm,CLAS:2019vsb} in panel (a), from Refs.~\cite{CLAS:2018xvc,JeffersonLabHallA:2020wrr,Li:2022fhh} in panel (b), and  from Ref.~\cite{CLAS:2018yvt} in panels (c) and (d). }\label{fig:exp}
\end{figure}

Figure~\ref{fig:exp} compares results on HMTs obtained using the momentum distribution parametrized as Eq.~(\ref{eq:momentum_distribution}) and the nucleon density distribution from the SHF model to the corresponding experimental data. Both the scaling factor $a_2(A)$, obtained from Eq.~(\ref{a2}) with the HMT fraction in deuteron of about 4\%~\cite{Passchier:2001uc}, and the $p_{np}/p_{pp}$ ratio, are within the error bar extracted from electron-nucleus scattering experiments, as shown in Figs.~\ref{fig:exp} (a) and \ref{fig:exp}  (b), respectively. In addition, our model catches the main feature of $np$ pairing in the HMT, by reproducing $\tilde{N}_n/\tilde{N}_p \sim 1$ for high-momentum nucleons with increasing $N/Z$ in Fig.~\ref{fig:exp} (c), and increasing relative high-momentum fractions for protons with increasing $N/Z$ in Fig.~\ref{fig:exp} (d). The consistency of the results from our model with the experimental data provides a solid baseline for the study of HMT with relativistic heavy-ion collisions.

{\it Measuring HMT with high-energy collisions.}
While it is still challenging to incorporate consistently the SRC effect into transport simulations at the mean-field level, efforts have been made in extracting the HMT of colliding nuclei through bremsstrahlung $\gamma$-ray emission in low-energy heavy-ion collisions~\cite{Xu:2025mvv}. In this study, we propose to measure the HMT of nucleons from spectator nucleons at forward rapidities in heavy-ion collisions at RHIC and LHC energy, free from the complicated midrapidity dynamics. Compared to nuclear reactions at low energies, the interaction between spectators and participants is negligibly small. Spectator neutrons can be directly measured by ZDC array~\cite{PHENIX:2000owy,ALICE:2013hur,Chu:2002}. Spectator protons will be deflected by the magnetic field in the beam optics, but are still measurable with specially designed detectors at forward rapidities~\cite{Tarafdar:2014oua} (see experiments by NA49 ~\cite{NA49:1998ocx} at SPS as well as those by ALICE~\cite{ALICE:2020bta,ALICE:1999edx,Alme:2010ke,ALICE:2022iqi} and ATLAS~\cite{ATLASnote:2020} at LHC). While the Fermi motion of nucleons in colliding nuclei in the beam direction is negligible compared to the beam energy, that in the transverse direction perpendicular to the beam direction is projected into a certain spatial distribution measured by ZDC array after collisions. The transverse momentum $k_T$ of spectator neutrons is linearly proportional to the transverse distance $r_\perp$ from the colliding point
\begin{align}
k_T &\approx r_\perp \times \frac{\sqrt{s^{}_{NN}/4-m^2}}{l},
\label{k_T}
\end{align}
where $m$ is the nucleon mass, $\sqrt{s^{}_{NN}}=200$ GeV represents the center-of-mass energy per nucleon pair, and $l=18$ m is the longitudinal distance from the colliding point to the ZDC. For the case of spectator protons, $l$ is longer due to the bent trajectory following the beam optics. The ZDC array has the dimension of about 10 cm width and 18 cm height at RHIC~\cite{Adler:2000bd}, and can achieve transverse spatial resolutions of roughly \(\sigma_{r_\perp} \sim 1 - 2\,~\mathrm{ cm}\), corresponding to the transverse momentum resolution of about $\sigma_{k_T} \sim 50 - 100$ MeV/c at top RHIC energy~\cite{STAR:2005btp,Crawford:2003}. 

We utilize the similar framework developed in previous studies~\cite{Liu:2022kvz,Liu:2022xlm} to extract information about free spectator nucleons in central relativistic heavy-ion collisions. The spatial coordinates of nucleons in colliding nuclei are sampled according to the density distribution obtained from SHF calculations, while their momenta are then sampled according to Eq.~(\ref{eq:momentum_distribution}).
A Glauber model with an NN scattering cross section of 42 mb at the top RHIC energy is used to identify spectator nucleons, which do not undergo any scatterings. The Wigner function approach~\cite{Chen:2003ava, Sun:2017ooe}, with the root-mean-square radii of light nuclei taken from Refs.~\cite{Ropke:2008qk, Tanihata:1985psr}, is then used to calculate the probability of these spectator nucleons to form clusters. While we consider the formation of light nuclei as heavy as $^6$Li and $^6$He, we have verified that the resulting distribution of free spectator nucleons, which do not form clusters, remains unchanged if we consider a maximum mass number of 3, 4, or 6 for clusters. The transverse momenta $k_T$ of free spectator nucleons are then used to construct the transverse spatial distribution according to Eq.~(\ref{k_T}), which can be measured by ZDC array. Nucleon production from the deexcitation of spectator matter from, e.g., electromagnetic excitation~\cite{Zhao:2022dac,Liu:2022xlm}, is less important, since it does not contribute to high-$k_T$ free spectator nucleons as a probe of HMT to be discussed later. 

\begin{figure}[htb]
    \centering
    \includegraphics[width=1\linewidth]{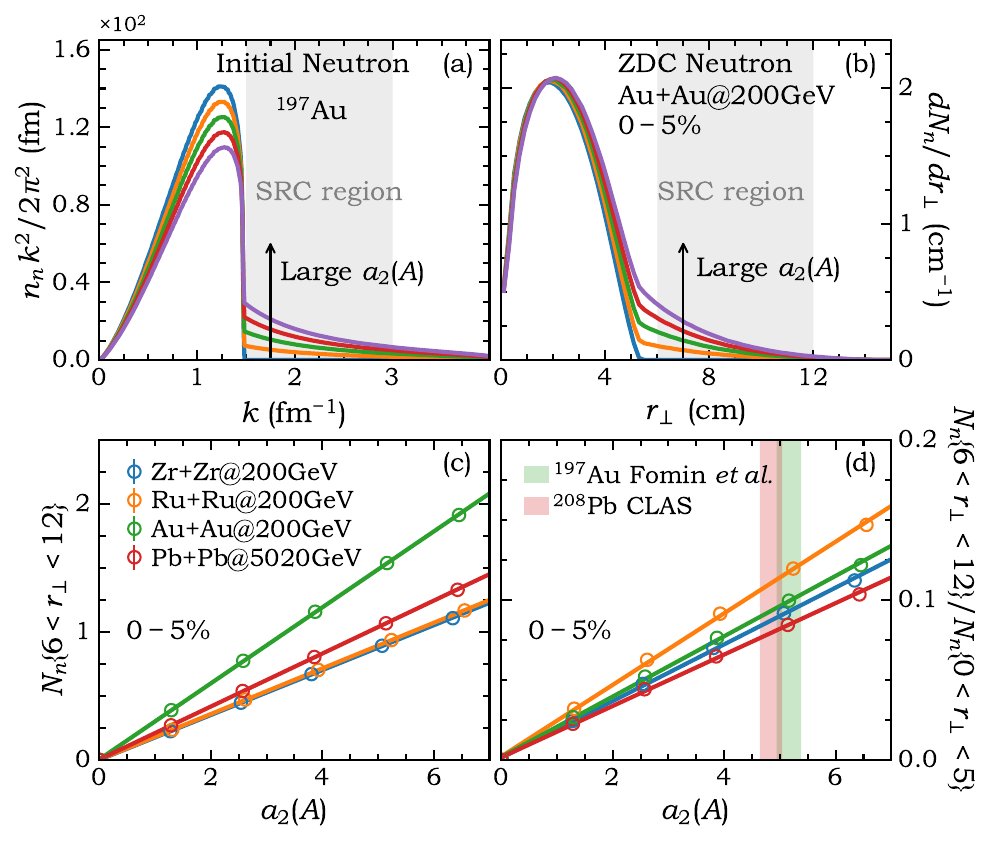}\\
    \vspace{-0.5cm}
    \caption{Upper: Momentum distributions of initial neutrons in $^{197}$Au (a) and transverse distance distributions of free spectator neutrons in central $^{197}$Au+$^{197}$Au collisions at $\sqrt{s^{}_{NN}}=200$ GeV (b) with different HMT fractions; Lower:  Linear correlations between the free spectator neutron yield within $6 < r_{\perp} < 12\ \mathrm{cm}$ (c) as well as the corresponding yield ratio with respect to that within $0 < r_{\perp} < 5\ \mathrm{cm}$ (d) and the scaling factor $a_2(A)$ of the colliding nuclei in different collision systems. Bands in panel (d) represent experimental data from electron-nucleus scatterings~\cite{Fomin:2011ng,Hen:2012fm,CLAS:2019vsb}. Results for $\sqrt{s^{}_{NN}}=5.02$ TeV are scaled to those at $\sqrt{s^{}_{NN}}=200$ GeV. }
    \label{fig:a2}
\end{figure}

{\it Results and Discussions.}
We start by displaying the neutron momentum distributions in $^{197}$Au in Fig.~\ref{fig:a2} (a). It is seen that the momentum distribution contains the low-momentum part and the high-momentum part, after the integration according to Eq.~(\ref{nk}), while the fraction of the HMT can be effectively adjusted by changing the value of $C$ in Eq.~(\ref{eq:momentum_distribution}). A larger HMT fraction leads to more energetic spectator nucleons propagating in the transverse direction, and thus more free spectator neutrons at large transverse distance $r_\perp$ measured by ZDC array, as shown in Fig.~\ref{fig:a2} (b). Therefore, measuring the number $N_n$ of ZDC neutrons at large $r_\perp$, e.g., $6 < r_{\perp} < 12\ \mathrm{cm}$, may help to reveal the HMT fraction in colliding nuclei. Actually, the neutron number at large $r_\perp$ is proportional to the HMT fraction as well as the scaling factor $a_2(A)$ of colliding nuclei in different collision systems, as shown in Fig.~\ref{fig:a2} (c). Since there are uncertainties to measure the absolute yield in experimental analysis, we propose to measure the ratio of ZDC neutron yields at large ($6 < r_{\perp} < 12\ \mathrm{cm}$) to small ($0 < r_{\perp} < 5\ \mathrm{cm}$) transverse distances, as shown in Fig.~\ref{fig:a2} (d), which is also proportional to $a_2(A)$ of colliding nuclei in different collision systems, characterizing the strength of SRCs.

\begin{figure}[htb]
\centering
\includegraphics[width=1\linewidth]{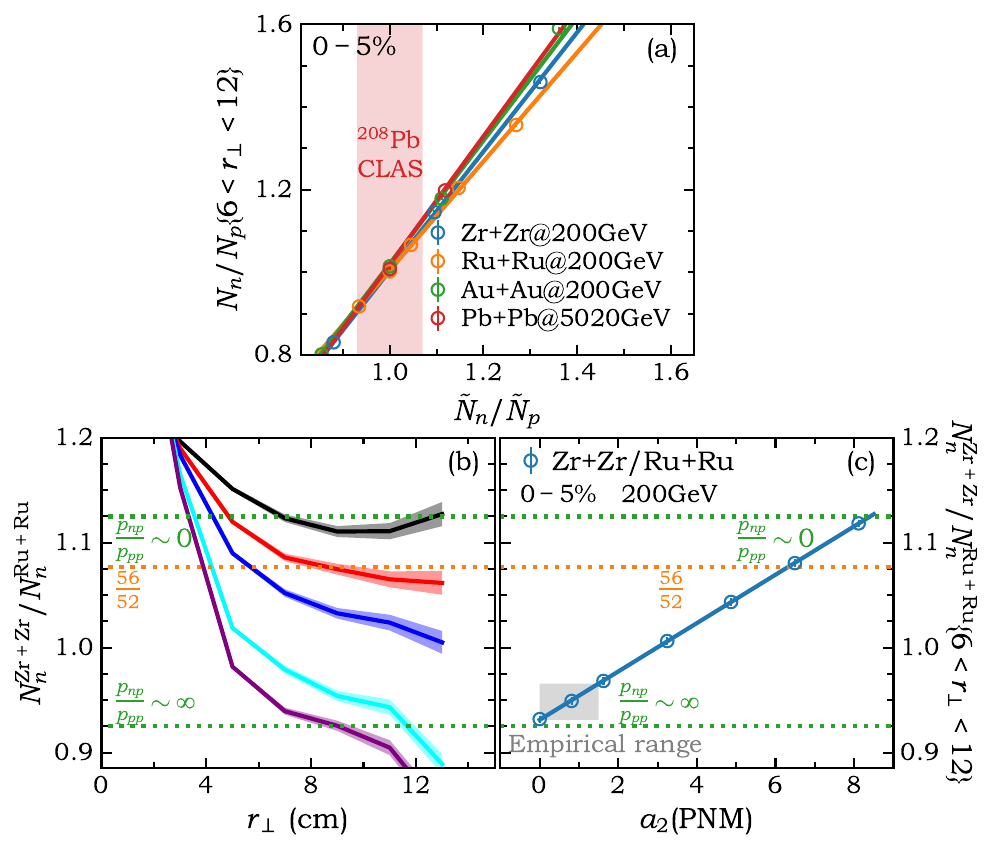}
\vspace{-0.8cm}
\caption{(a): Linear correlation between the free spectator neutron-to-proton yield ratio at large transverse distances ($6 < r_{\perp} < 12~\mathrm{cm}$) and the neutron-to-proton number ratio in the HMT of colliding nuclei, with the band representing the electron-nucleus scattering data by CLAS Collaboration taken from Ref.~\cite{CLAS:2018yvt}. (b) Ratios of the free spectator neutron transverse distance distribution in central $^{96}\mathrm{Zr}+^{96}\mathrm{Zr}$ collisions to that in central $^{96}\mathrm{Ru}+^{96}\mathrm{Ru}$ collisions from different $\alpha_1$.
(c): Linear correlation between the ratio of the free spectator neutron yield at large transverse distances ($6 < r_{\perp} < 12~\mathrm{cm}$) in central $^{96}\mathrm{Zr}+^{96}\mathrm{Zr}$ collisions to that in central $^{96}\mathrm{Ru}+^{96}\mathrm{Ru}$ collisions and the scaling factor $a_2$ of pure neutron matter (PNM). 
}
\label{fig:isospin}
\end{figure}


We further illustrate that measuring ZDC nucleons at large $r_\perp$ may reveal the isospin dependence of HMT and SRCs. By adjusting the value of $\alpha_1$ in Eq.~(\ref{eq:momentum_distribution}), we are able to change the relative neutron and proton fraction in the HMT, related to $p_{np}/p_{pp}$ in Fig.~\ref{fig:exp} (b). As discussed above, the yields of ZDC neutrons and protons at large $r_\perp$ originate from those in the HMT of colliding nuclei. Therefore, measuring the neutron-to-proton yield ratio at large $r_\perp$ can help to extract the neutron-to-proton number ratio in the HMT, and the corresponding linear relation is shown in Fig.~\ref{fig:isospin} (a) for different collision systems. Using this extracted neutron-to-proton number ratio, one can further determine the $np$ dominance in short-range correlations via Eq.~\eqref{eq:npppratio}. Experimentally, it is difficult measure spectator protons and select those at effective $r_\perp$ as the same as spectator neutrons with proper detecting efficiency corrections. Here we further propose to measure the ratio of the free spectator neutron yield at large $r_\perp$ in central $^{96}\mathrm{Zr}+^{96}\mathrm{Zr}$ collisions to that in central $^{96}\mathrm{Ru}+^{96}\mathrm{Ru}$ collisions to extract the isospin dependence of SRCs. The advantage of choosing isobaric collision systems is that the colliding nuclei share the same factor $f_{\mathrm{sr}}(A)$ in Eq.~(\ref{eq:SRCpairNum}). Figure~\ref{fig:isospin} (b) illustrates the ratio of the free spectator neutron $r_\perp$ distribution in central $^{96}\mathrm{Zr}+^{96}\mathrm{Zr}$ to $^{96}\mathrm{Ru}+^{96}\mathrm{Ru}$ collisions. With a more negative $\alpha_1$, the ratio decreases more rapidly with increasing $r_\perp$, demonstrating a smaller neutron number in the HMT in more neutron-rich $^{96}\mathrm{Zr}$. Figure~\ref{fig:isospin} (c) displays a linear relation between the proposed ratio at large $r_\perp$ and the scaling factor $a_2(\mathrm{PNM})$ for pure neutron matter at the saturation density, with the value of the latter adjusted also by $\alpha_1$ in Eq.~(\ref{eq:momentum_distribution}) and serving as a representative quantity related to the isospin dependence of SRCs. The upper and lower limits of the ratio are
\begin{equation}
\frac{\tilde{N}_{n}^\mathrm{Zr}}{\tilde{N}_{n}^\mathrm{Ru}} = \frac{\left[N\cdot (N-1)\right]_\mathrm{Zr}}{\left[N\cdot (N-1)\right]_\mathrm{Ru}}>1
\end{equation}
and
\begin{equation}
\frac{\tilde{N}_{n}^\mathrm{Zr}}{\tilde{N}_{n}^\mathrm{Ru}} = \frac{\left[N \cdot Z\right]_\mathrm{Zr}}{\left[N \cdot Z\right]_\mathrm{Ru}}<1,
\end{equation}
corresponding to $p_{np}/p_{pp}=0$ and $p_{np}/p_{pp}=\infty$, respectively, as indicated in Fig.~\ref{fig:isospin} (b) and (c). While the ratio of neutron number in $^{96}$Zr to $^{96}$Ru is 56/52, the proposed ratio is expected to be smaller using the default parametrization from the empirical $a_2(\mathrm{PNM})$, since there are less $np$ pairs in neutron-rich $^{96}$Zr than in neutron-deficient $^{96}$Ru.  

{\it Summary and Outlook.}
We propose to measure free spectator neutrons at large transverse distances detectable by ZDC array in relativistic heavy-ion collisions, which serve as a probe of the HMT in colliding nuclei induced by SRCs, free from the complicated dynamics at midrapidity and with less FSIs. To account for the uncertainties in experimental measurements, we propose to measure the neutron yield ratio at large to small transverse distances to extract the strength of SRCs, and the neutron yield ratio at large transverse distances in central $^{96}\mathrm{Zr}+^{96}\mathrm{Zr}$ collisions to that in central $^{96}\mathrm{Ru}+^{96}\mathrm{Ru}$ collisions to extract the isospin dependence of SRCs. Similar method can be potentially generalized to $d$+$^{197}$Au, $^3$He+$^{197}$Au, $^{16}$O+$^{16}$O, and $^{20}$Ne+$^{20}$Ne collisions in order to extract the HMT and SRCs in light nuclei. The proposed measurement can hopefully be carried out by investigating existing ZDC data in previous relativistic heavy-ion collision experiments. Our proposal serves as an alternative probe of measuring HMT and SRCs in nuclei compared to the traditional electron-nucleus scattering experiments, toward an accurate measurement of nucleus structure and nuclear force.

{\it Acknowledgments.}
We acknowledge helpful discussions with Bao-An Li. This work is supported by the National Key Research and Development Program of China under Grants Nos. 2023YFA1606701 and 2022YFA1604900, the National Natural Science Foundation of China under Grant Nos. 12375125, 12225502, and 12147101, the China Postdoctoral Science Foundation under Grant No. 2024M750489, the Natural Science Foundation of Shanghai under Grant No. 23JC1400200, and the Fundamental Research Funds for the Central Universities.

\bibliography{ref}


\end{document}